# Multiple beam steering using dynamic zone plates on a micro-mirror array.

**David M. Benton**

Aston University, Aston Triangle, Birmingham, B4 7ET.

E-mail: d.benton@aston.ac.uk

**Abstract** Three dimensional laser beam steering has been demonstrated using a single optical device – a DMD micro-mirror array. Laser beam focus position is controlled using dynamically adjustable zone plates. These zone plates take the form of elliptical Fresnel zone plates or other variations such as binary Gabor zone plates. Active beam control can be realised without the need for a pair of galvanometer mirrors and a focusing lens. Writing multiple zone plate patterns to the DMD enables multiple focussed spots to be generated and controlled independently.

## Introduction

One of the most useful properties of lasers is the ability to generate well directed, low divergence beams. This enables laser beam based interactions such as sampling or communications at specific remote locations. In the majority of cases where active control of the laser pointing is required a set of galvanometer mirrors is used but these have a tendency to be large, heavy, power hungry and expensive. The size and weight also limits the response times. Non-mechanical methods of beam steering are then attractive [1]1 and phased array type approaches have been investigated for this purpose [2][3]. This can show potentially rapid steering with no moving parts.

The DMD micro-mirror array by Texas Instruments is a MEMS device that has been commercially available since 1996 and its predominant application is that of image projection. There are however many other applications for which the DMD has been used [4] including spectroscopy [5], switching [6], single pixel remote sensing [7] and laser glare suppression [8] to name just a few. The DMD is a binary amplitude spatial light modulator (SLM) composed of an array of individually addressable tilting micro mirrors. In comparison with phase modulating SLMs the DMD has certain advantages including a fast response rate (22kHz) and wide spectral range of operation (UV – IR). The spatial modulation control of a DMD has been used to shape the properties of a reflected beam [9], produce Laguerre-Gaussian beams [10] and non-diffracting Bessel beams [11]. Very little attention has been paid to using DMDs for beam steering applications, perhaps due to the off axis reflection and potential efficiency issues. Refai et al [12] constructed a beam scanning system by placing the DMD at the focal plane of a lens and switching on pixels to act as point sources. When off axis these sources were collimated at an angle to the optical axis and an output beam could be steered by changing the required pixels. However this system only used a small fraction of the available pixels and light at any one time. This also produced beams steered in 2 dimensions. Recently Cheng et al [13] used a DMD to provide rapid axial focussing in two photon microscopy.

The Fresnel zone plate (FZP) has been understood for a very long time and its focussing ability has been used to concentrate fluxes of radiation where lenses are impractical due to size or material constraints e.g. x-rays or neutrons. The FZP is a binary amplitude device and therefore a natural choice for use with a DMD. It is perhaps more fashionable to refer the FZP as a binary curved wavefront hologram but in this paper the traditional term is used.

In this paper 3 dimensional non-mechanical beam steering is observed, using the simplest of optical systems, by writing FZPs to a DMD device. The paper progresses as follows: A brief discussion of the FZP is followed by modelling of the FZP on a DMD system. Correction of astigmatism is achieved followed by the observation of real time beam steering with control over x and y position and longitudinal beam focus. Multiple beams are produced and finally a more unusual zone plate pattern is used. This paper restricts itself to discussion and observation of three dimensional steering of a beam, not to more involved matters of wavefront phase analysis or optimal focussing regimes, this will be the subject of further work.

**The Fresnel zone plate**

The FZP is a series of concentric alternating transparent and opaque rings that act to focus light via diffraction. Opaque regions prevent light of one polarity of phase from the reaching the focal point. The radius of the *n*th ring from the centre of the FZP is given by[14]

$$r_n = \sqrt{nf\lambda + \frac{n^2\lambda^2}{4}} \cong \sqrt{nf\lambda}$$

Where *f* is the focal length of the FZP and $\lambda$ the wavelength of the incident radiation. The DMD has pixel sizes of 13.6μm – A FZP lens with a focal length of 1m, with a wavelength of 600nm would have first zone radius of 0.77mm, equivalent to 56.9 pixels and easily achievable. Thus a DMD can directly form optical structures with lab bench scale relevance. The FZP has been shown to offer a resolution limit better than a lens depending on the number of zones used and whether positive or negative zones are used [15][16]. FZPs are normally used in transmission or planar reflection, their use with a pixelated array of titled mirrors requires consideration.

**Modelling of an off-axis tilted mirror zone plate**

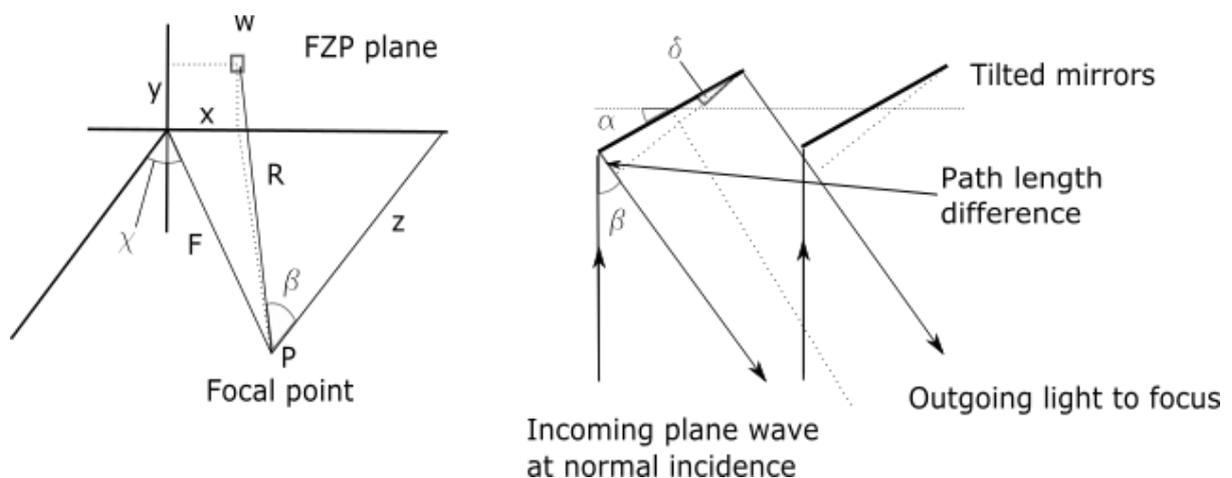

*Figure 1. Geometry for a micro mirror array focussing light at a point P at an angle to the optical axis.*

The diagrams in Figure 1 represent a DMD array in the FZP plane, focussing light via reflection of an incoming plane wave, at a point P. Light that is normally incident upon the mirror array is diverted by the tilt of each mirror in a fashion that is similar to a blazed diffraction grating. If the tilt of each mirror is an angle α then coherent light forms bright beams governed by interference conditions where the optical path difference at an angle χ is an integer number of wavelengths, close to the angle 2α. This is considered to be the axis from the centre of the pattern to the focal point P. The Fresnel zone plate works by only summing together waves that have the same polarity of phase – that is opaque regions prevent phase contributions that would reduce the net intensity. We therefore need to determine the optical path difference from each point on the micro mirror array to the off axis focal point in order to determine the pixel state. The angle χ is determined from

$$\sin(\chi) = \frac{n\lambda}{w} \qquad 2$$

Where $n$ is an integer, $\lambda$ is the incident wavelength and $w$ is the dimension of an individual mirror.

The optical path difference for each pixel position $O(x,y)$ is determined by:

$$O(x,y) = F - R(x,y) \qquad 3$$

Where F is the focal length of the zone plate and R(x,y) is the distance from pixel location (x,y) to the focal point P. The distance is given by:

$$R(x,y) = \sqrt{(F\sin(\chi) - x)^2 + y^2 + (F\cos(\chi))^2} \qquad 4$$

The location x=0, y=0 is assumed to be the centre of the pattern where the condition for integer n is true. Hence the phase contribution from each pixel at point P is

$$\phi(x,y) = O(x,y)\frac{2\pi}{\lambda} \qquad 5$$

This gives the phase difference from the average position of the pixel and varies rapidly in the x direction due to the off axis view of the zone plate. We must also consider the effect of path difference across each mirror. At the centre of the pattern this is $n\lambda$ but varies as the angle made between pixel and P changes with x position across the array.

At this point it is worth reflecting on why this needs investigating. Consecutive zones within the zone plate represent path length differences of around half a wavelength. When the DMD mirrors are tilted at 12°, the path length difference *across each mirror* is multiple wavelengths – typically > 9 for wavelengths around 635nm. Thus it is not obvious that a FZP pattern should work at all with such a device.

The optical path difference between rays at each end of a mirror is

$$OM = w\sin(\alpha) - w\sin(\delta) \qquad 6$$

Where δ is the angle between the mirror and a line orthogonal to the outgoing rays. Through geometry we find the relation

$$\delta = \beta - \alpha \qquad 7$$

Where α is the tilt angle of each mirror and β is the angle made between incoming and outgoing light towards the focal point P and is dependent upon mirror position.

There is also a very slight angular difference across each mirror for the outgoing rays which are directed at a common focal point. This will also be dependent on the mirror position, however given that F>>w, this effect shows negligible change across the array in comparison to the change in angle β and can be neglected.

The angle β can be obtained from

$$\beta(x) = \arctan\left[\frac{f\sin(\chi) - x}{f\cos(\chi)}\right] \qquad 8$$

Thus the optical phase difference across each mirror is then

$$\phi_m(x.y) = w\left(\sin(\alpha) - \sin(\beta(x) - \alpha)\right)\frac{2\pi}{\lambda} \qquad 9$$

The Fresnel zone plate is then produced by switching the mirror state S(x,y) on (0) or off (1) according to the generating principle similar to that expressed in [16]

$$S(x,y) = 0 \;\; if \;\; \sin\left(\phi(x,y) + \phi_m(x.y)\right) > 0$$
$$S(x,y) = 1 \;\; if \;\; \sin\left(\phi(x,y) + \phi_m(x.y)\right) \leq 0 \qquad 10$$

Using the values for the DMD where w=13.6μm, α=12° (a consequential diffraction order of n=9), and a minimum pixel dimension of 760 a zone plate with a focal length F=35cm for an illuminating wavelength of 635nm is shown in Figure 2. Given the off axis reflected nature of this zone plate it is still very circular. Cross sections through the vertical and horizontal directions show a slight asymmetry.

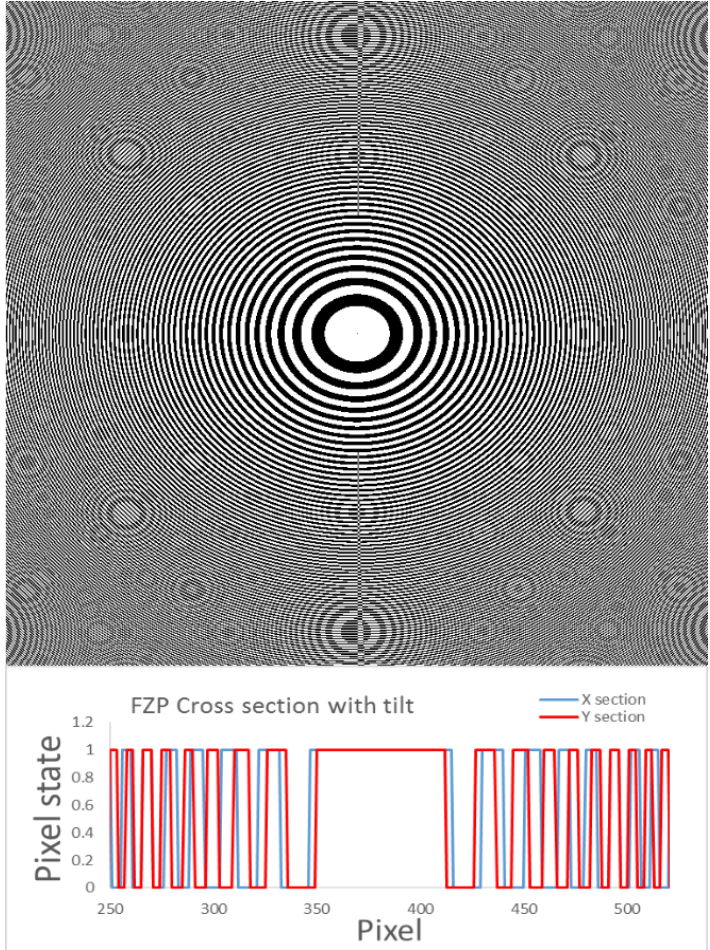

*Figure 2. Modelled zone plate for off axis focussing with a micro mirror array (top). The lower plots are of the cross section through the centre in both x and y directions.*

The focal length of the zone plate is determined by the radius of the first zone according to equation 1. This gives an expected radius of $4.36 \times 10^{-4}$ m, or 32 pixels. The values of the first zone radius from the modelled zone plate are 31 pixels in the x direction and 34 pixels in the y direction, in reasonable agreement with the calculation for an on axis zone plate.

The axis to point P has been defined from the centre of the pattern, thus if an offset to the pattern centre is provided, the focal point P can be moved relative to the central diffraction order position. The pixel phase can then be written as

$$\phi(x,y) = O(x - x_0, y - y_0)\frac{2\pi}{\lambda} \qquad 11$$

where $x_0$ and $y_0$ are offsets. Thus the focal spot location can be controlled in 3 dimensions by defining $x_0$, $y_0$ and $r_1$ for the zone plate.

## Zone plates on the Micro mirror array

A DLP discovery 4100 development kit (Vialux) containing a DMD with 1024 x 768 pixels and a pixel spacing of 13.6μm was used in this work. The control circuitry and software routines enable the pattern on the DMD to be rapidly swept through a series of predetermined frames at a controlled rate. The frames are pre-calculated and sent to the DMD control circuitry and pattern buffers via a USB connection. This is capable of modulating individual pixels at rates up to 22kHz depending on the grayscale addressing mode that is used. However for realtime interactive control of a laser beam position a single frame was sent to the DMD and updated with a new pattern as required. This inevitably reduces the response rate of the system due to lags resulting from frame calculation and USB update rates.

A simple experimental arrangement was used to observe focussed spot behaviour, where the DMD is the only optical component involved in the focussed spot production. Light from a fibre coupled laser diode (635nm) was collimated at a beam size larger than the DMD device, and incident on the micro mirror array at normal incidence. A USB CCD camera was placed adjacent to the DMD to observe the focussed beam from the DMD array appearing on a screen. The pattern on the DMD was controlled using a LabView program which generated zone plate patterns based upon the conventional formula given the required focal length.

Given the similarity between on-axis zone plates and tilted mirror off-axis zone plates it is acceptable to use the conventional equations for zone plate generation in order to write zone plates to the micro mirror array. Illuminating the array with normal incidence plane wave produces an off axis array of diffraction spots, this being the Fraunhofer diffraction pattern of a 2 dimensional array of source mirrors. Each diffraction order contains a focussed spot arising from the zone plate pattern imposed onto the array.

**Focussing efficiency**

The DMD is a 2 dimensional blazed diffraction grating and a comprehensive derivation of the diffraction effects can be seen in [9][17]. Here we will take a generic approach from which the intensity profile $I(x,y)$ at the mirror can be represented as a convolution of multiple 2D functions where $x$ and $y$ are coordinates in the array plane

$$I(x,y) = rect(mirror\ size) * comb(mirror\ spacing) * rect(array\ size) * F(x,y)$$

Where the rect(mrror size) is a rectangular function representing the 2D shape extent of each individual mirror, comb(mirror spacing) is a 2D repeating function representing the array spacing structure, rect(array size) puts rectangular shaped boundaries on the array and F(x,y) is the FZP pattern written across the array. The Fraunhofer diffraction pattern produced is the Fourier transform of this intensity profile with $u$ and $v$ representing the coordinates in the image plane which is generically

$$I'(u,v) = \left\{ sinc\left(\frac{1}{mirror\ size}\right) . comb\left(\frac{1}{mirror\ spacing}\right) . sinc\left(\frac{1}{array\ size}\right) . F'(u,v) \right\}^2$$

Where F'(u,v) is the Fourier transform of the FZP pattern. This pattern can be seen in Figure 3 where focal spots resulting from the FZP pattern *(F'(u,v))* can be seen within each diffraction order (comb function) whose intensity is modulated by the sinc(1/mirror size) function. Distributing the incident intensity across multiple orders clearly reduces the intensity in any resultant focussed spot. The focussed intensity in the principle orders can be calculated using the following:

$$I = I_0\ f_A\ f_{split}\ f_e\ f_{off}\ f_{tilt}$$

Where $I_0$ is the incident intensity, $f_A$ is the fractional power incident upon the rectangular array, $f_{split}$ is the fraction directed in the positive tilt direction, $f_e$ is the focussing efficiency of the zone plate into the primary first order focus, $f_{off}$ is the reduction factor due to the order being offset from the centre of the diffraction pattern and $f_{tilt}$ is the area reduction of the mirrors due to being tilted relative to the input beam.

The fractional beam power is calculated from the overlap of a circular Gaussian beam of diameter 22mm, with a HWHM of 5mm overlapping the rectangular area of the DMD. The power fraction was calculated by simulation and found to give $f_A$=0.85. The input power is split 50% into the positive and negative tilt directions because the sum area of 'transmitting' and 'opaque zones' is the same. The focussing efficiency into the first order focus of a FZP is $f_e=1/\pi^2$ [16]. The reduction factor $f_{off}$ is determined from the value of the sinc(1/mirror size) function. The angular spacing of the comb function is a=$\lambda$/d , and assuming the comb orders are equally spaced about the centre of the diffraction pattern, the diffraction intensity is

$$f_{off} = \left\{sinc\left(\frac{d}{\lambda}\frac{a}{2}\pi\right)\right\}^2 = 0.4$$

Finally the $f_{tilt}$ fraction represents the reduction in area due to the mirror being tilted with $f_{tilt}$ =cos(12°). The input intensity was 1.1mW which leads to an expected focal spot intensity of 17µW, in good agreement with the measured values and with values from [17](where all micro mirrors are used). Thus efficiency of directing light into one of the principal orders is around 1% for this wavelength where the orders are not at the maximum diffraction efficiency. When the FZP pattern is translated to steer the beam to the centre of the diffraction pattern the intensity increased to 42µW, an efficiency of 3.8%.

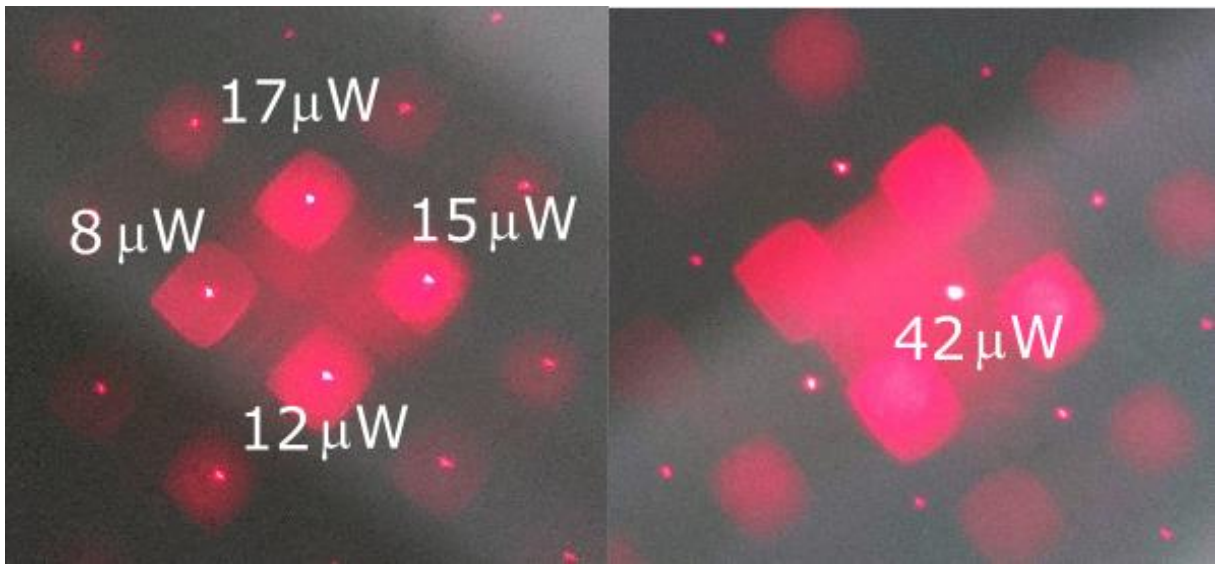

*Figure 3. The intensity pattern at the focal plane with a central FZP (left image) and an offset FZP (right image)*

## Astigmatism

At longitudinal positions away from the focus the beam shows clear signs of astigmatism. The zone plate tilt relative to the focal point P makes the zone plate appear to be elliptical rather than circular, compressed in the x direction. The zone plate then appears to have a different focal length in the x and y directions. This can be overcome by scaling the x position by a factor $1/\cos(2\alpha)$, where $2\alpha$ is referred to as the obliquity, making the zone plate appear circular from position P. In addition this now elliptical zone plate must be rotated azimuthally by 45° because the micro mirrors pivot along a diagonal axis at 45° to the pixel axes.

Figure 4 shows images of focal spots at a fixed imaging plane where the zone plate focal length is varied. The two columns of focal spots compare the effect of obliquity =0° (circular) on the left and obliquity =24°(elliptical) on the right, showing a clear reduction in astigmatism for the elliptical zone plate.

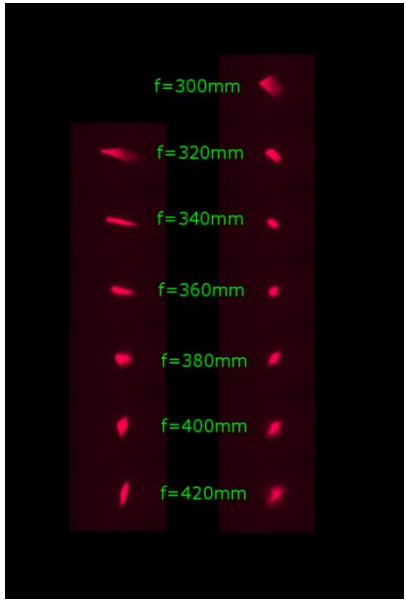

*Figure 4. Focal spots at a fixed position as the zone plate focal length is varied. Spots on the left are a basic zone plate, spots on the right are elliptical plates to correct for astigmatism.*

Data captured from the camera was used to fit a Gaussian function to the focused spot intensity distribution and enabled the beam spot size and position to be determined as zone plate parameters were varied. These can be seen in Figure 5 where the top plot shows the spot size Gaussian width for an elliptical zone plate with obliquity=24°, as the zone plate focal length is varied. The central plot shows how the amount of obliquity affects the spot size and the lower plot shows the focal spot position moving laterally as the zone plate pattern is offset from the centre of the DMD.

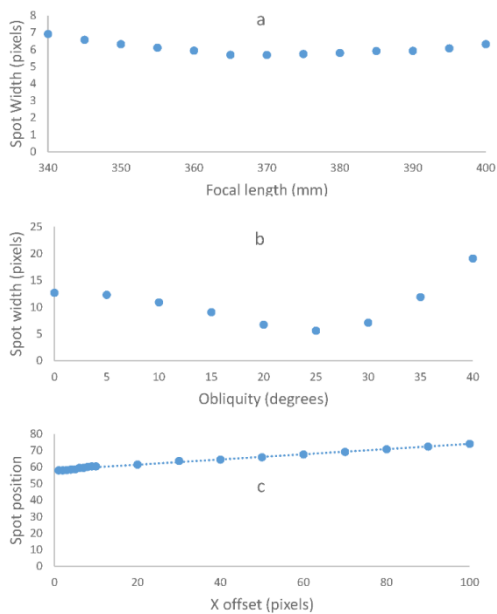

*Figure 5. Variations in focal spot parameters as zone plate parameters are varied. The top plot (a) shows the spot width variation at a fixed axial distance with zone plate focal length. Plot (b) shows spot width variation with zone plate obliquity and plot(c) shows spot centroid displacement with pattern offset.*

**Dynamic beam steering**

The camera image was used to control beam position by clicking on the live camera image with a mouse and dragging the beam spot around the screen. The mouse position provided a scaled value for zone plate offsets. Figure 6 shows images of a focussed beam spot and the zone plate patterns used to create them. The spot can be seen separated from the unaffected diffraction order as a consequence of the zone plate pattern offset. Whilst this system operated in real time, there was an issue with the computation of zone plates for every frame slowing down the response. There are a number of ways of improving the processing time but an effective way was found to be calculating an initial zone plate with a given focal length, in which the pattern is larger, typically 5X larger than the micro mirror array itself. Obtaining the zone plate pattern for any particular offset required simply the selection of the correct sub array of the correct size. Changes in focal length required recalculation of the zone plate pattern.

A focussed laser beam spot can be seen at each diffraction order with typically 4 bright orders close to double the mirror tilt angle. Many orders can be observed in both x and y directions but the majority of the intensity is within these few orders [17]. The angular spacing between orders has been calculated and measured at 2.67°. When steered to a point halfway between orders contiguous steering can be achieved by continuing across orders thereby increasing the effective steering range to several multiples of the angular diffraction order spacing. The steering of the focussed spot is being achieved by beam offsetting rather than by angle steering. At large offsets the zone width becomes smaller than the resolution of the mirrors the zones become ineffective. The expression for the zone plate focal length in terms of zone width $\Delta r_n$ is

$$f = \frac{2\, r_n\, \Delta r_n}{\lambda}$$

By setting the zone width equal to the mirror size it can be seen that there is a fixed relationship between the zone plate focal length and the nth zone radius at which the mirror resolution is reached. The displacement of the focussed spot from the zero offset position at the resolution limit is $r_n(f)$ and this makes an angle

$$\tan(\theta) = \frac{offset}{f} = \frac{r_n(f) - D/2}{f}$$

where *D* is dimension of the DMD array. At long focal lengths (>1m) this angle tends to 1.2° and is smaller at shorter wavelengths. Thus the zone plate resolution cannot be realised at the pattern edges when significant offsets are applied. Nevertheless beam steering across orders has been observed and this is most likely to result from the zone plate behaving as a curved diffraction grating [19] Aliasing and zone truncation occurs even before the zone width reaches the mirror size.

In addition an autofocussing routine was created which measured the spot size both in terms of area and in terms of x and y widths. This routine sought to take advantage of the ability of the dynamic array to change between widely different focal lengths without the need to progress through all intermediate focal lengths. Spot sizes from a set of 3 trial focal lengths were recorded where an initial spacing between settings was used. From these values a polynomial function was used to predict the expected point of best focus. The process was then repeated around this expected focal length with a reduced spacing between trial values and stopped when a minimum spot size was achieved. This could typically find a focus in nine steps but requires more development and should be fully reported later.

This system was constructed to show that information from a camera image can be used to direct a beam towards a chosen point. This could, in principle be from an image processing algorithm that could be tracking a target and maintaining a beam focussed on that target. The images within Figure 6 were produced by dragging the image of the focus spot seen using a mouse pointer in the camera image and calculating the necessary offset pattern.

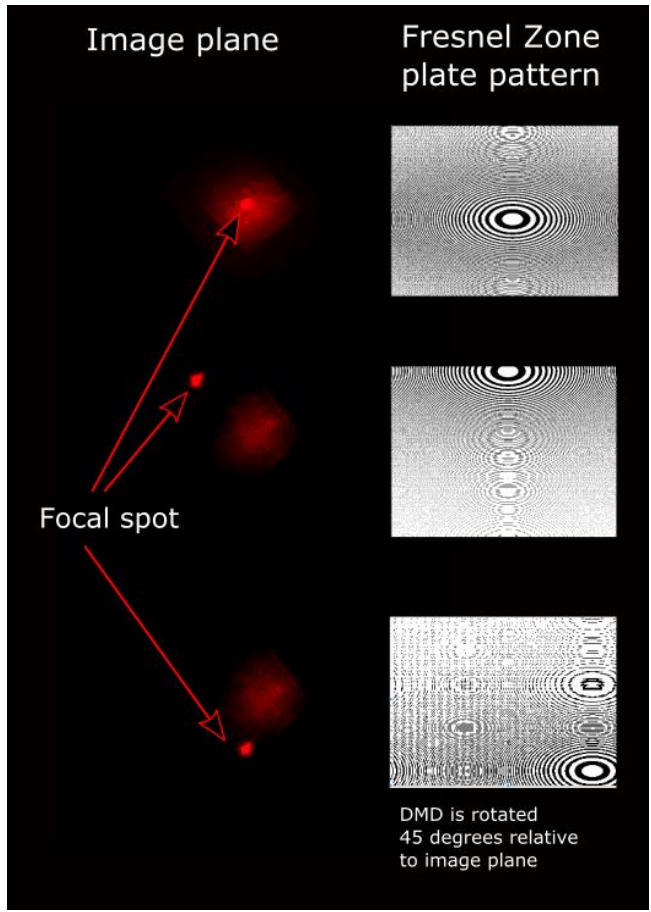

*Figure 6. Images of focal spots being offset from the diffraction order, with accompanying zone plate patterns.*

By overlaying separate zone plate patterns multiple focussed beams can be generated as can be seen in Figure 7. In this case 2 identical zone plates were offset from each other and a sequence of steps made where the beams orbit around a central point, causing the focussed beams to circulate around a central point. The patterns were ORed together such that points with negative phase contributions from each do not interfere with the other. This of course affects the efficiency.

Each pattern can in principle be controlled independently such that output beams have differing focal lengths and could track independent targets. This is a direction for further development.

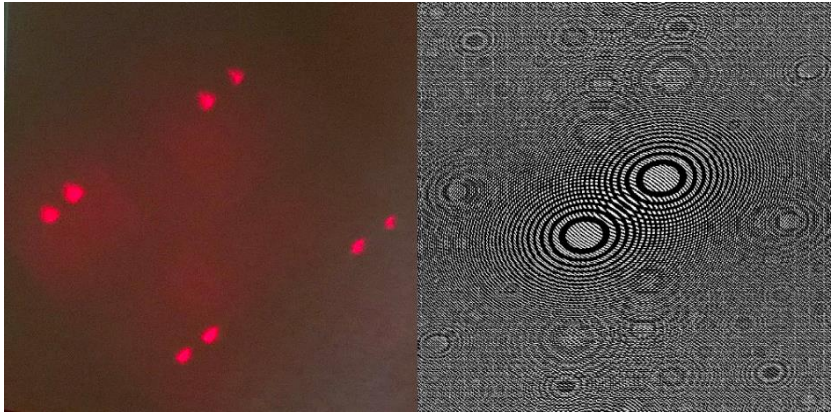

*Figure 7. Dual focused spots formed at each diffraction order using the dual zone plate pattern on the left.*

## Binary Gabor zone plates

The Fresnel zone plate is a binary version of the grayscale Gabor zone plate and differs by having multiple points of longitudinal focus, where the Gabor plate has a single focus. Beynon[20] described a binary version of the Gabor zone plate. At any radius r the value of the transmission is

$T(r) = 1 + \cos\left(\left(\frac{r}{r_1}\right)^2\right)$ and this is true at any azimuthal angle. However so long as the circular integral around $2\pi$ amounts to this value, how this value is arrived at is less important. Thus by blocking out an arc of the circle equivalent to the value *(1-T(r))* the equivalent Gabor plate transmission can be obtained with binary values. Beynon implemented this by dividing the circle into sectors and dividing each sector in the ratio *T(r):1-T(r)*. Binary Gabor zone plates were implemented on the micro mirror array and were subjected to the same elliptical distortion and rotation as the Fresnel zone plates. Focussed laser spots and the generating binary Gabor zone plate are shown in Figure 8. The binary Gabor zone plate can act as a direct replacement for the Fresnel zone plate in all applications and hence works equally well in 3 three dimensional beam steering applications.

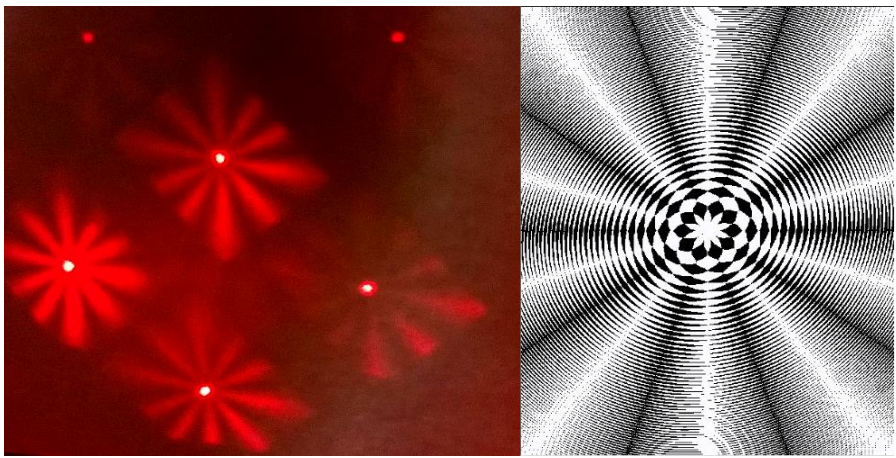

*Figure 8. Focal spots generated from a binary Gabor zone plate pattern with 10 sectors shown on the right.*

## Conclusion

The Fresnel zone plate and a binary amplitude spatial light modulator in the form of a micro mirror array are a natural combination. Using real-time adjustable zone plate parameters it is possible to perform 3-dimensional beam steering where the beam focus (axial z-position) is controlled by adjusting the size of the zone plate pattern (first zone radius $r_1$) and x and y location of the beam spot is controlled by offsetting the centre of the zone plate pattern.   This results in the use of single planar device  - the DMD mirror array – performing the functions conventionally requiring a pair of scanning mirrors and a variable focus lens system. Not only does this represent a reduction in the size and weight in comparison to the conventional approach, but allows a different functionality. Both the scanning mirrors and focussing lens are serial in nature whereas the micro-mirror array represents a random access approach where any point within the field of view can be accessed within a single step. Further to this multiple beams can be controlled independently. In this work it has been shown that the DMD array can be used in conjunction with a co-located camera to allow active beam pointing and focussing to be performed and that binary zone plates are an effective means of achieving this.

**References.**

[1]. Oh, C. W., Tangdiongga, E., & Koonen, A. M. J. (2014). Steerable pencil beams for multi-Gbps indoor optical wireless communication, *39*(18), 5427–5430.
[2]. McManamon, P. F., Bos, P. J., Escuti, M. J., Heikenfeld, J., Serati, S., Xie, H., & Watson, E. a. (2009). A review of phased array steering for narrow-band electrooptical systems. *Proceedings of the IEEE*, *97*(6), 1078–1096. doi:10.1109/JPROC.2009.2017218
[3]. Van Acoleyen, K., Komorowska, K., Bogaerts, W., & Baets, R. (2011). One-dimensional off-chip beam steering and shaping using optical phased arrays on silicon-on-insulator. *Journal of Lightwave Technology*, *29*(23), 3500–3505. doi:10.1109/JLT.2011.2171477.
[4]. Dudley, D., Duncan, W. M., & Slaughter, J. (2003). <Emerging digital micromirror device (DMD) applications. *Proc. SPIE 4985, MOEMS Display and Imaging Systems, 14 (January 20, 2003);*, (Dmd), 14–25. doi:10.1117/12.480761.
[5]. Graff, D. L., & Love, S. P. (2013). Real-time matched-filter imaging for chemical detection using a DMD-based programmable filter, *8618*, 86180F. doi:10.1117/12.2002694
[6]. Blanche, P.-A., Carothers, D., Wissinger, J., & Peyghambarian, N. (2013). DMD as a diffractive reconfigurable optical switch for telecommunication, *8618*, 86180N. doi:10.1117/12.2006428
[7]. Ma, J. (2009). Single-Pixel Remote Sensing. *IEEE Geoscience and Remote Sensing Letters*, *6*(2), 199–203. doi:10.1109/LGRS.2008.2010959
[8]. Ritt, G., & Eberle, B. (2015). Automatic Laser Glare Suppression in Electro-Optical Sensors. *Sensors*, *15*(1), 792–802. doi:10.3390/s150100792
[9]. Ren, Y. X., Lu, R. De, & Gong, L. (2015). Tailoring light with a digital micromirror device. *Annalen Der Physik*, *527*(7-8), 447–470. doi:10.1002/andp.201500111.
[10]. Lerner, V., Shwa, D., Drori, Y., & Katz, N. (2012). Shaping Laguerre–Gaussian laser modes with binary gratings using a digital micromirror device. *Optics Letters*, *37*(23), 4826–8. doi:10.1364/OL.37.004826
[11]. Gong, L., Ren, Y.-X., Xue, G.-S., Wang, Q.-C., Zhou, J.-H., Zhong, M.-C., Li, Y.-M. (2013). Generation of nondiffracting Bessel beam using digital micromirror device. *Applied Optics*, *52*, 4566–75. doi:10.1364/AO.52.004566
[12]. Refai, H. H., Sluss, J. J., & Tull, M. P. (2007). Digital micromirror device for optical scanning applications. *Optical Engineering*, *46*(8), 085401. doi:10.1117/1.2768978

[13]. Cheng, J., Gu, C., Zhang, D., Wang, D., & Chen, S.-C. (2016). Ultrafast axial scanning for two-photon microscopy via a digital micromirror device and binary holography. *Optics Letters*, *41*(7), 1451–1454. doi:10.1364/OL.41.001451

[14]. Young, M. (1972). Zone Plates and Their Aberrations. *Josa*, *62*(8), 972–976. doi:10.1364/JOSA.62.000972

[15]. Stigliani, D. J. J., Mittra, R., & Semonin, R. G. (1967). Resolving Power of a Zone Plate. *Journal of the Optical Society of America*, *57*, 610. doi:10.1364/JOSA.57.000610

[16]. Cao, Q., & Jahns, J. (2004). Comprehensive focusing analysis of various Fresnel zone plates. *Journal of the Optical Society of America. A, Optics, Image Science, and Vision*, *21*(4), 561–71.

[17]. Ma, J. (2012). Magnification of optical image in holography projection using lensless Fresnel holography. *Optical Engineering*, *51*(8), 85801. http://doi.org/10.1117/1.OE.51.8.085801

[18]. Park, M.-C., Lee, B.-R., Son, J.-Y., & Chernyshov, O. (2015). Properties of DMDs for holographic displays. *Journal of Modern Optics*, *62*(19), 1600–1607. doi:10.1080/09500340.2015.1054445

[19]. Blanchard, P. M., & Greenaway, A. H. (1999). Simultaneous multiplane imaging with a distorted diffraction grating. *Applied Optics*, *38*(32), 6692–9.

[20]. Beynon, T. D., Kirk, I., & Mathews, T. R. (1992). Gabor zone plate with binary transmittance values. Optics Letters, 17(7), 544–6.